\documentclass[12pt]{article}

\usepackage{newtxtext,newtxmath}
\usepackage[T1]{fontenc}

\usepackage{graphicx}

\usepackage[letterpaper,margin=1in]{geometry}

\renewenvironment{abstract}
	{\quotation}
	{\endquotation}

\date{}

\makeatletter
\renewcommand{\fnum@figure}{\textbf{Figure \thefigure}}
\renewcommand{\fnum@table}{\textbf{Table \thetable}}
\makeatother

\usepackage{scicite}

\usepackage{url}

\def\scititle{
	Tunable superconducting diode effect in a topological nano-SQUID
}
\title{\bfseries \boldmath \scititle}

\author{
	Ella~Nikodem$^{1\dagger}$,
    Jakob~Schluck$^{1\dagger}$,
	Max~Geier$^{2}$,
    Micha\l ~Papaj$^{3}$,
	Henry~F.~Legg$^{4,5}$,\and
    Junya~Feng$^{1}$,
    Mahasweta~Bagchi$^{1}$,
    Liang~Fu$^{2}$,
    Yoichi~Ando$^{1\ast}$\and
	\small$^{1}$Physics Institute II, University of Cologne, Zülpicher Str. 77, 50937 Köln, Germany.\and
	\small$^{2}$Department of Physics, Massachusetts Institute of Technology, Cambridge MA 02139, USA.\and
    \small$^{3}$Department of Physics and Texas Center for Superconductivity at the University of Houston (TcSUH),\and \small Houston, TX 77204, USA.\and
    \small$^{4}$Department of Physics, University of Basel, Klingelbergstrasse 82, 4056 Basel, Switzerland.\and
    \small$^{5}$SUPA, School of Physics and Astronomy, University of St Andrews,\and \small North Haugh, St Andrews, KY16 9SS, United Kingdom.\and
	\small$^\ast$Corresponding author. Email: ando@ph2.uni-koeln.de\and
	\small$^\dagger$These authors contributed equally to this work.
}

\begin{document} 

% Insert the title and author list
\maketitle

\begin{flushleft}
{\bf Short title:}\\
Topological nano-SQUID superconducting diode
\end{flushleft}

\begin{flushleft}
{\bf Teaser:}
A large tunable superconducting diode effect is found in a side-contacted topological-insulator nanowire Josephson junction.
\end{flushleft}

% Abstract, in bold
% There are strict length limits, and not all formats have abstracts.
% Consult the journal instructions to authors for details.
% Do not cite any references in the abstract.
\begin{abstract} \bfseries \boldmath
A Josephson diode passes current with zero resistance in one direction but is resistive in the other direction. While such an effect has been observed in several platforms, a large and tunable Josephson diode effect has been rare. Here we report that a simple device consisting of a topological-insulator (TI) nanowire side-contacted by superconductors to form a lateral Josephson junction presents a large diode effect with the efficiency $\eta$ reaching 0.3 when a parallel magnetic field $B_{||}$ is applied. Interestingly, the sign and the magnitude of $\eta$ is tunable not only by $B_{||}$ but also by the back-gate voltage. This diode effect can be understood by modeling the system as a nano-SQUID, in which the top and bottom surfaces of the TI nanowire each form a line junction and $B_{||}$ creates a magnetic flux to thread the SQUID loop. This model further shows that the observed diode effect marks the emergence of topological superconductivity in TI-nanowire-based Josephson junction. 
\end{abstract}

% The first paragraph of any Science paper does NOT have a heading
% Nor is it indented
\noindent

\section{Introduction}
%\vspace{-5mm}

%Nonreciprocal superconducting transport effects that lead to a 
Superconducting (SC) diode effects have become a topic of current interest \cite{Ando2020,Nadeem2023}, mainly because they allow for the realization of useful SC circuit components for  low-dissipation electronics with potential applications in quantum computing \cite{Frattini2017May,Grimm2020Aug} and superconducting neural networks \cite{Schegolev2022}. Both bulk superconductors and Josephson junctions may present a diode effect when inversion and time-reversal symmetries are broken, and the latter is called Josephson diode effect \cite{Nadeem2023, Wu2022}. 
A wide range of mechanisms can give rise to a SC diode effect, including asymmetric vortex pinning in a bulk SC slab \cite{Vodolazov2005, Vandevondel2005,Golod2022Jun}, Meissner screening  \cite{Hou2023Jul}, asymmetric current-phase relationship in a junction \cite{Fulton1972,Souto2022,Ciaccia2023,Lu2023}, magnetochiral anisotropy  \cite{Ando2020,He2022May,Baumgartner2022}, finite-momentum pairing  \cite{Yuan2022, Davydova2022, Pal2022,Nakamura2024Mar}, and topological phase transitions  \cite{Banerjee2023d, Legg2023, Cayao2024}.

%The mechanisms underlying the diode effect can be relatively trivial, like asymmetric vortex pinning in a bulk SC slab \cite{Vodolazov2005, Vandevondel2005} or asymmetric current-phase relationship in a junction \cite{Fulton1972}, but it can also be highly nontrivial and related to exotic phenomena like finite-momentum pairing \cite{Yuan2022, Davydova2022, Pal2022} or a topological phase transition \cite{Banerjee2023, Legg2023, Cayao2024}.

The Josephson diode effect is of particular interest because of its potential for high controllability \cite{Lu2023}. One way to achieve a highly tunable device with a high diode efficiency is to make an asymmetric Superconducting Quantum Interference Device (SQUID) involving two or more Josephson junctions  \cite{Souto2022, Ciaccia2023}. We have recently discovered \cite{Nikodem2024} that a single Josephson junction consisting of a topological-insulator (TI) nanowire side-contacted by two SC electrodes on both sides presents pronounced oscillations of the critical current $I_c$ as a function of the magnetic field $B_{||}$ applied parallel to the nanowire axis, resulting in an $I_c(B_{||})$ behavior akin to a SQUID. 
Importantly, the period in $B_{||}$ corresponds to the magnetic flux $\Phi_0 = h/2e$, the SC flux quantum, threading the TI nanowire. These results strongly suggest that the side-contacted TI-nanowire junction forms an intrinsic nano-SQUID, where the top and bottom surfaces of the nanowire effectively work as two parallel SC-normal-SC (SNS) junctions, while the bulk is insulating; the threading magnetic flux $\Phi$ imposes a SC phase difference between the two SNS junctions, leading to the $I_c$ oscillations. 
%In this device, the asymmetry between top- and bottom junction is controllable by a gate voltage.
Therefore, it is interesting to determine if such a nano-SQUID formed intrinsically in a TI-nanowire junction would present a Josephson diode effect, as in other asymmetric SQUIDs \cite{Fulton1972, Souto2022, Ciaccia2023}. 
If it does, such a diode has the potential advantage of being highly tunable by both the parallel magnetic field $B_{\parallel}$ and the back-gate voltage $V_G$, which introduce time-reversal and inversion symmetry breaking, respectively.

In this paper, we demonstrate that the side-contacted TI-nanowire junction indeed presents a large and tunable Josephson diode effect. Intriguingly, the diode efficiency $\eta$ changes sign with both $B_{\parallel}$ and the back-gate voltage $V_G$, and its magnitude reaches 0.3 which is 
comparable to the largest reported for single Josephson junctions  \cite{Baumgartner2022} without vortex trapping. 
%among the largest reported for a Josephson diode. 
Our devices are based on TI nanowires with 1 -- 2 $\upmu$m length, 10 -- 20 nm thickness, and 60 -- 80 nm width, dry-etched from an exfoliated flake of the bulk-insulating TI material BiSbTeSe$_2$ and side-contacted by Nb electrodes. This structure allows for efficient back-gating, which tunes the asymmetry between the two SQUID arms, an essential ingredient of the Josephson diode effect in the SQUID configuration \cite{Souto2022}. The interface between the TI and Nb has been reported to be sufficiently transparent to give rise to a highly skewed current-phase relationship \cite{Kayyalha2020}, which is also crucial for the Josephson diode effect \cite{Souto2022}; in fact, the spin-momentum locking in the TI surface state guarantees a perfectly-transmitted Andreev mode 
 \cite{Tkachov2013, Schluck2024} and enhances the skewness \cite{Kayyalha2020}. Our full three-dimensional simulations of the TI nanowire side-contacted by a conventional superconductor reproduce all the key features observed experimentally, giving confidence in the nano-SQUID origin of the Josephson diode effect. Moreover, our theoretical analysis shows that the sign change in $\eta$ corresponds to a topological phase transition in the TI nanowire, which should be accompanied by the emergence of Majorana zero-modes  \cite{Nikodem2024}. Hence, the relatively simple platform reported here is interesting not only for future SC electronics but also for Majorana physics.

\section{Results}

\subsection{Josephson diode effect in a side-contacted TI nanowire junction}

\begin{figure}
\centering
\includegraphics[width=\textwidth]{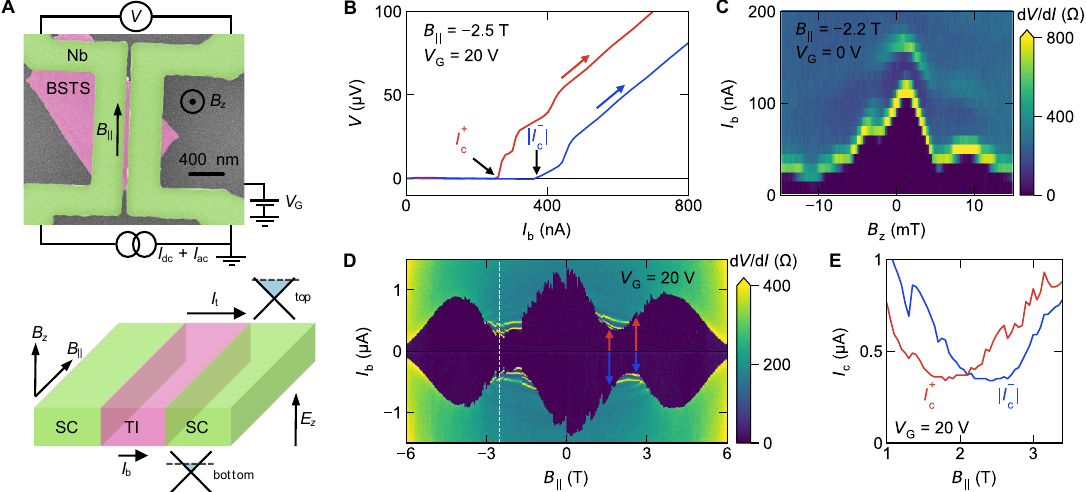}
\caption{\linespread{1.05}\selectfont{}
{\bf Device schematics and the behavior of the critical current.} 
(\textbf{A}) False-color SEM image of device A with schematics of the pseudo-four-terminal measurement. An exfoliated TI flake is dry-etched into a nanowire (pink). The etched area is filled with Nb (green) to form a sandwich junction. There are remaining TI pieces that do not play any role. The lower drawing shows that the supercurrents flow through the top and bottom topological surface states, whose chemical potentials can be made different by a back-gate creating the electric field $E_z$; a magnetic field applied parallel to the nanowire axis causes the Josephson diode effect.
(\textbf{B}) $I$-$V$ characteristics of device A taken at $B_{||}=-2.5$ T and $V_\mathrm{G} = 20$ V. For better comparability with the positive bias curve (red), the negative bias curve (blue) was flipped horizontally and vertically. The current sweep was always from zero as indicated by the red and blue arrows. 
(\textbf{C}) Color map of d$V$/d$I$ as function of out-of-plane magnetic field $B_z$ and $I_\mathrm{b}$ at $B_{||}=2.2$ T and $V_\mathrm{G} = 0$ V in device A. 
(\textbf{D}) Color map of d$V$/d$I$ as function of $B_{||}$ and $I_\mathrm{b}$ at $V_\mathrm{G} = 20$ V in device A. Red and blue arrows highlight the diode effect. The white dashed line mark the $B_{||}$ value where the data in \textbf{B} was taken. 
(\textbf{E}) Plot of $I_\mathrm{c}^+$ and $|I_\mathrm{c}^-|$ extracted from the data in \textbf{D} as function of $B_{||}$ near an $I_c$-minimum at $B_{||} \approx$ 2 T; here, the critical currents are approximated by the $I_b$ value at which $dV/dI$ exceeded 50 ohms. 
}
\label{fig:Fig1}
\end{figure}

Figure \ref{fig:Fig1}A shows a false-color scanning-electron microscope (SEM) image of our device A, including the schematics of the pseudo-four-terminal measurement. 
%The device is fabricated on a degenerately-doped Si substrate coated with a 290-nm-thick SiO$_2$ dielectric layer, through which the back-gate voltage $V_G$ is applied. The TI nanowire is proximitized by 
The Nb electrodes had the transition temperature $T_\mathrm{c}\approx$ 7 K, and the in-plane upper critical field $H_\mathrm{c2, \parallel}$ was slightly above 6 T, which was the highest magnetic field available in this experiment. As described in Ref.  \citenum{Nikodem2024}, the precise alignment of the parallel magnetic field $B_{\parallel}$ using a vector magnet was crucial for this type of device. Even a small $B_{z}$ component will cause the ordinary Fraunhofer interference effect and reduce the $I_c$, messing up the $I_c(B_{\parallel})$ behavior. We show in Fig. \ref{fig:Fig1}C an example of the Fraunhofer pattern measured as a function of $B_{z}$ in the presence of $B_{\parallel} = 2.2$ T, where one can see that $B_{z}$ of just a few mT reduces $I_c$ substantially (the asymmetry in the pattern in Fig. \ref{fig:Fig1}C is due to an anomalous phase created by $B_{||}$ \cite{Assouline2019}, which does not open a gap but shifts the Fermi surface in the momentum space \cite{Yuan2018}). In the following, we focus on results obtained from device A (shown in Fig. \ref{fig:Fig1}A), and additional data from three more devices are shown in the supplement. Unless otherwise noted, the data were acquired at the base temperature of our dry dilution refrigerator ($\approx$30 mK).

In Fig. \ref{fig:Fig1}B, we show the $I$-$V$ characteristics measured at $B_{||}=-2.5$ T and $V_\mathrm{G}=20$ V to present the Josephson diode effect: The data for positive current $I^+$ are shown in red, while the data for negative current $I^-$ are shown in blue and are plotted for $|I^-|$ with flipped voltage.
%such that one can easily compare the two to see the nonreciprocity of the critical current.
In these measurements, we always swept the current from zero to avoid hysteresis or other complications. The data in Fig. \ref{fig:Fig1}B show the critical currents $I_\mathrm{c}^+ = 250$ nA and $|I_\mathrm{c}^-| = 390$ nA, giving the diode efficiency $\eta \equiv (I_\mathrm{c}^+-|I_\mathrm{c}^-|)/(I_\mathrm{c}^++|I_\mathrm{c}^-|) \approx -0.23$.
%this means that when the bias current $|I_\mathrm{b}|$ is the range of 250 -- 390 nA, the junction is superconducting in one direction and normal conducting in the other direction. Here, the efficiency $\eta$ of this diode effect defined as $\eta \equiv (I_\mathrm{c}^+-|I_\mathrm{c}^-|)/(I_\mathrm{c}^++|I_\mathrm{c}^-|)$ is about $-0.23$.
Figure \ref{fig:Fig1}D maps the differential resistance d$V$/d$I$ measured across the junction as a function of $B_{||}$ and the bias current $I_\mathrm{b}$ taken at $V_\mathrm{G}=20$ V; the dark area corresponds to zero resistance and its boundary marks $I_c$. The pronounced oscillations of $I_c$ as a function of $B_{||}$ having the periodicity of $\Phi_0$ ($= h/2e$) is due to the intrinsic nano-SQUID formed in our junction \cite{Nikodem2024}. 

One can see in Fig. \ref{fig:Fig1}D that for most of the $B_{||}$ values, $|I_c^+|$ and $|I_c^-|$ are not the same. This is particularly clear in the region near the $I_c$-minima (e.g. around $B_{||} \approx$ 2 T).
Interestingly, across the $I_c$-minima, the sign of $\eta$ changes, as highlighted by red and blue arrows. This sign change is more clearly demonstrated in Fig. \ref{fig:Fig1}E, where the $B_{||}$-dependencies of $I_c^+$ and $|I_c^-|$ are plotted in red and blue, respectively. The crossing of $I_c^+(B)$ and $|I_c^-(B)|$ at $B_{||} \approx 2$ T corresponds to the sign change in $\eta$.

\subsection{Gate-tunability of the Josephson diode effect}

\begin{figure}
\centering
\includegraphics{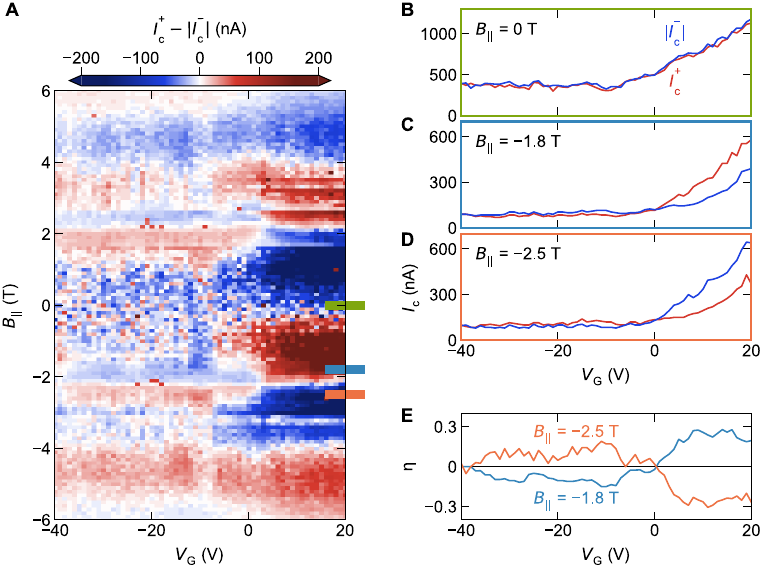}
\caption{\linespread{1.05}\selectfont{}
{\bf Tunability of the diode efficiency.} 
(\textbf{A}) Color mapping of $I_\mathrm{c}^+ - |I_\mathrm{c}^-|$ as a function of $V_\mathrm{G}$ and $B_{||}$ in device A, demonstrating the tunability of the Josephson diode effect. 
(\textbf{B-D}) Raw data of  $I_\mathrm{c}^+(V_G)$ and $|I_\mathrm{c}^-(V_G)|$ used for producing \textbf{A} at $B_{||}$ = 0, $-1.8$, and $-2.5$ T;
these $B_{||}$ values are marked in \textbf{A} with corresponding colors.
(\textbf{E}) $V_G$-dependence of the diode efficiency $\eta$ for $B_{||}$ = $-1.8$ T and $-2.5$ T calculated from the data in \textbf{C} and \textbf{D}. }
\label{fig:Fig2}
\end{figure}

In our TI-nanowire junction, one can tune the sign and the magnitude of the diode effect not only with $B_{||}$ but also with $V_\mathrm{G}$.
The gate-tunability of $I_\mathrm{c}$ in zero field is shown in Fig. \ref{fig:Fig2}B, where $I_\mathrm{c}$ changes roughly by a factor of three with $V_\mathrm{G}$.
In Fig. \ref{fig:Fig2}A we plot $I_\mathrm{c}^+-|I_\mathrm{c}^-|$ as a function of $B_{||}$ and $V_\mathrm{G}$ as a color map, where the white color corresponds to zero (i.e. no diode effect). One can see that the diode effect oscillates with $B_{||}$, resulting in alternation of red and blue; the magnitude of $I_\mathrm{c}^+-|I_\mathrm{c}^-|$ can exceed 200 nA at positive $V_G$, which is large since $|I_\mathrm{c}|\lesssim1$ \textmu A at all times (see Fig. \ref{fig:Fig1}D). Notice that $I_\mathrm{c}^+ - |I_\mathrm{c}^-|$ is essentially antisymmetric in $B_{||}$, as expected.

\begin{figure}
\centering
\includegraphics{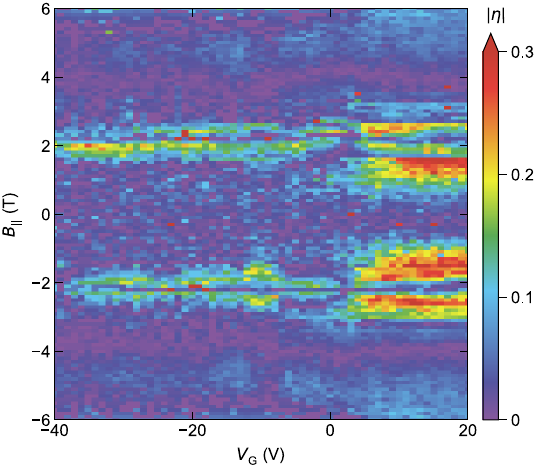}
\caption{\linespread{1.05}\selectfont{}
{\bf Magnitude of the diode efficiency.} 
Color mapping of $|\eta|$ as a function of $V_\mathrm{G}$ and $B_{||}$ in device A, showing that the magnitude of the diode efficiency can reach 0.3 near the sign-switching magnetic field $B_{||} \approx \pm$2 T. 
}
\label{fig:Fig3}
\end{figure}

Importantly, we observed that for $|B_{||}|\lesssim 3$ T and at $|B_{||}|\approx 6$ T, the nonreciprocity $I_\mathrm{c}^+-|I_\mathrm{c}^-|$ presents a gate-induced sign change. 
%across $V_\mathrm{G} \approx$ $-5$ - 0 V, where the diode effect nearly disappears.
To make this behavior clear, we plot the $V_G$-dependencies of $I_\mathrm{c}^+$ and $|I_\mathrm{c}^-|$ in red and blue lines, respectively, for $B_{||}$ = 0, $-1.8$, and $-2.5$ T in Figs. \ref{fig:Fig2}B-D. 
%The data for $B_{||}$ = 0 demonstrates the strong gate tunability of our device as indicated by more than a doubling of the critical current in the gate range studied.
One can see that the two lines cross at $-5$ V $\lesssim V_\mathrm{G} \lesssim$ 0 V in $B_{||}$ = $-1.8$ and $-2.5$ T, while they are essentially indistinguishable at 0 T.  One can also notice that the way the sign change occurs is opposite between $B_{||}$ = $-1.8$ and $-2.5$ T; namely, $\eta$ ($\propto I_\mathrm{c}^+-|I_\mathrm{c}^-|$) changes from negative to positive with increasing $V_G$ in $B_{||}$ = $-1.8$ T, while it is opposite in $B_{||}$ = $-2.5$ T. 
%The $B_{||}$ values corresponding to Figs. \ref{fig:Fig2}B-D are indicated by colored triangles in Fig. 2A.
To further visualize the $V_G$-induced sign change, in Fig. \ref{fig:Fig2}E we plot $\eta$ in $B_{||}=-1.8$ and $-2.5$ T as a function of $V_G$, where the peculiar sign-changing behavior is apparent. Note that $|\eta|$ reaches 0.3 which is large \cite{Nadeem2023, Banerjee2023, Cayao2024, Lu2023, Souto2022, Ciaccia2023}. Figure \ref{fig:Fig3} depicts the parameter regime where this large diode effect is observed; one can see that $|\eta| \approx$ 0.3 is achieved near $B_{||} \approx \pm$2 T where the sign change of $|\eta|$ takes place.
%It is useful to see that $\eta$ is nearly zero in the $V_G$ range from $-5$ to 0 V, which seems to suggest that inversion symmetry breaking caused by $V_G$ is necessary for the diode effect to take place, which is reasonable given that an asymmetry in the nano-SQUID is a prerequisite for the nonreciprocity.

\begin{figure}
\centering
\includegraphics{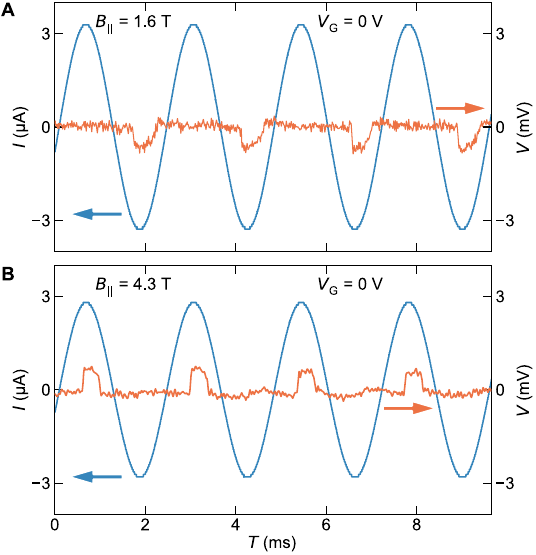}
\caption{\linespread{1.05}\selectfont{}
{\bf Visualization of the rectification effect. } 
(\textbf{A} and \textbf{B}) Time-resolved measurement of the voltage drop across the Josephson junction (orange curve) in response to a sinusoidal AC current with an amplitude of $\sim$3 $\upmu$A (blue), highlighting the rectification effect. The data in \textbf{A} and \textbf{B} were acquired at $B_{||}$ = 1.6 and 4.3 T, respectively, where the sign of $\eta$ was opposite. The measurement was performed on device B. }
\label{fig:Fig4}
\end{figure}

To demonstrate the rectification behavior resulting from our Josephson diode, we performed simple time-domain measurements using another junction, device B: We recorded the voltage waveform when a simple sinusoidal AC current was applied to the junction. 
%For $V_G$ = 0 V, this device showed a positive $\eta$ in $B_{||}= 1.6$ T, while $\eta$ was negative in $B_{||}= 4.3$ T.
In $B_{||}= 1.6$ T, as shown in Fig. \ref{fig:Fig4}A, when the AC-current amplitude is $\sim$3 $\upmu$A (which is larger than $|I_\mathrm{c}^-|$ but
smaller than $|I_\mathrm{c}^+|$), the junction presents a rectified waveform with only negative voltage appearing for negative current.  In $B_{||}= 4.3$ T, on the other hand, the sign change in $\eta$ results in the opposite rectification as shown in Fig. \ref{fig:Fig4}B. This experiment demonstrates that the diode polarity is easily switched by the magnetic field. One can similarly switch the diode polarity with gate voltage, as shown in Fig. \ref{fig:Fig2}E. The AC response in Fig. \ref{fig:Fig4} is limited by the low-pass filters in the signal line; it would be interesting to elucidate the intrinsic frequency limit of this type of SC diode.

\subsection{Theoretical analysis and simulations}

The behavior of the Josephson junction can be modeled phenomenologically as a nano-SQUID where the supercurrent is carried separately by the top and bottom surfaces with skewed current-phase relations,
\begin{equation}
I_{\rho}(\theta_\rho)=I_{0,\rho}\left[\sin\left(\theta_\rho \right)+S_{\rho}\sin\left(2 \theta_\rho \right)\right]\,,\label{eq:CPR_rho}
\end{equation}
where $I_{0,\rho}$ and $S_{\rho}$ determine the magnitude and the skewness
of the current-phase relation, 
and $\theta_\rho$ is the superconducting phase difference at the top and bottom surfaces ($\rho={\rm t},{\rm b}$). 
The total Josephson current is $I(\theta)=\sum_{\rho={\rm t},{\rm b}}I_{\rho}(\theta_\rho)$.
The magnetic flux $\Phi$ threading through the TI nanowire dictates $\theta_{\rm t} - \theta_{\rm b} = 2\pi \frac{\Phi}{\Phi_0} \equiv \phi$.
%; we choose $\theta_{\rm t} = \theta + \phi$, $\theta_{\rm b} = \theta$ as a convenient parametrization. 
The area that determines
$\Phi$ in the nano-SQUID model is affected by the 
localization depth of the TI surface states below the nominal surface  \cite{Nikodem2024},
as well as the London penetration depth of the superconductor. 
The sign of $S_\rho$ determines the skewness direction. 
Skewed current-phase relation arise naturally in high-transparency
Josephson junctions and are right skewed ($S_{\rho}<0$) \cite{Beenakker1992,Golubov2004Apr}, and intermediate transparency leads to $|S_{\rho}|\ll1$; a direct measurement of the current-phase relation for the same TI material was reported in Ref.~ \citenum{Kayyalha2020}, which found a typical skewness $|S_{\rho}| \approx 0.2$.

A Josephson diode effect in our TI-based nano-SQUID occurs under
the following conditions: (i) A magnetic flux through the nanowire
breaking time-reversal symmetry, (ii) an asymmetry between the top and
bottom surfaces, $I_{0,{\rm t}}\neq I_{{\rm 0,b}}$ and/or $S_{{\rm t}}\neq S_{{\rm b}}$,
which breaks inversion symmetry~ \cite{Legg2021,Legg2022,Legg2022MCA}, and (iii) a finite skewness of the current-phase
relation $|S_{\rho}|>0$. Time-reversal and inversion symmetry must
be broken because their action reverts the direction of current flow.
The gate-induced top/bottom asymmetry is enough for the present diode effect and there is no need for bulk nor structure inversion asymmetry.
The skewness $|S_{\rho}|>0$ ensures the presence of higher harmonics
in the current-phase relation that are necessary for the critical
currents in opposite directions to be distinct.
Conversely, for zero skewness both individual and total current-phase relations are sinusoidal so that in this case the critical currents in both directions are equal. While the nonreciprocal transport and a SC diode effect have both been discussed for TI nanowires with the current flowing along the wire axis \cite{Legg2022, Legg2022MCA}, in the present nano-SQUID setup the current runs perpendicular to the wire axis.

We further performed a full three-dimensional numerical simulation of a TI-nanowire Josephson junction using a tight-binding model (details presented in Materials and Methods). The comparison between the simulation results and the relevant experimental data for the critical currents $I_\mathrm{c}^+$ and $|I_\mathrm{c}^-|$, together with the diode efficiency, is shown in Fig.~\ref{fig:Fig6}. As the chemical potential is placed within the bulk band gap of the 3D TI, the current is carried only through the top and bottom surfaces of the nanowire, which yields the nano-SQUID picture. 
The simulation faithfully reproduces the behavior of critical currents as well as the magnitude of the diode effect, exhibiting sign changes of $\eta$ at $\Phi=\pm \Phi_{0}/2$. Crucially, the diode effect appears in the calculation only when the asymmetry between the top and bottom surfaces is present.

In the nano-SQUID model, time-reversal symmetry is restored at integer multiples of half a flux quantum, $\Phi=\frac{n}{2}\Phi_{0},\ n\in\mathbb{Z}$. As a consequence, the critical current asymmetry must flip when the flux is mirrored around each half flux quantum, $I_{{\rm c}\pm}(\frac{n}{2}\Phi_{0}+\delta\Phi)=-I_{{\rm c}\mp}(\frac{n}{2}\Phi_{0}-\delta\Phi)$. This requires the diode efficiency $\eta$ to be periodic and odd around each $\Phi=\frac{n}{2}\Phi_{0}$. 
This behavior is approximately observed in the microscopic simulation and in the experiments. Note that the deviation from the expected periodicity near $\Phi/\Phi_0 = \pm 1$ is due to additional effects such as the suppression of superconductivity in the Nb leads and Zeeman coupling.
To complement this microscopic simulation, the behavior of the diode effect was also calculated from the phenomenological nano-SQUID model [Eq.~\eqref{eq:CPR_rho}], reproducing the key features of the experimental data, as shown in Fig.~S7 of the supplement.

%Fig. \ref{fig:Fig5} shows critical currents $I_{{\rm c}\pm}$ and diode efficiency $\eta=\frac{I_{{\rm c+}}-|I_{{\rm c-}}|}{I_{{\rm c+}}+|I_{{\rm c-}}|}$ calculated from Eq. \ref{eq:CPR_rho} with asymmetry $I_{{\rm 0,t}}/I_{{\rm 0,b}}=1.2$ and equal skewness $S=S_{{\rm t}}=S_{{\rm b}}=-0.2$. The critical current is minimal around odd integer multiples of half a flux quantum $\Phi=\frac{n}{2}\Phi_{0},\ n\in\mathbb{Z}$ by canceling contributions from top and bottom surface. Precisely at the integer multiples of half a flux quantum, $\Phi=\frac{n}{2}\Phi_{0},\ n\in\mathbb{Z}$, time-reversal symmetry in the phenomenological model is restored. As a consequence, the critical current asymmetry must exchange when the flux is mirrored around each half flux quantum, $I_{{\rm c}\pm}(\frac{n}{2}\Phi_{0}+\delta\Phi)=-I_{{\rm c}\mp}(\frac{n}{2}\Phi_{0}-\delta\Phi)$. This requires the diode efficiency $\eta$ to be periodic and odd around each $\Phi=\frac{n}{2}\Phi_{0}$. 

In a closed system at equilibrium, the superconducting phase difference across the junction must satisfy the condition that  current is conserved:
\[
I_{{\rm t}}(\theta_{0} + \phi)=-I_{{\rm b}}(\theta_{0})\ .
\]
Here, a gauge is chosen to include the effect of the magnetic flux in the phase difference of the top junction (i.e. $\phi = 2\pi\Phi/\Phi_0$).
This equation has two or more solutions due to the periodicity of the supercurrent with the superconducting phase difference, but the system selects the solution that minimizes the free energy of
the junction, fixing the equilibrium phase bias $\theta_0$. At zero temperature, the free energy is given by the Josephson energy $E_{J}=\frac{\hbar}{2e}\int^{\theta}d\theta'I(\theta')$.
As the flux is increased from $\Phi = 0$ to $\Phi_0$, the gauge-invariant phase difference $\gamma_{t,b} = \theta_{t,b} + e \int \mathbf{A} d\mathbf{l}  /\hbar $ on the weaker junction winds from $0$ to $2\pi$ while the phase difference of the other junction remains close to $0$.
For small skewness or large asymmetry, this phase winding is continuous. For larger skewness above a critical value set by the asymmetry, the equilibrium phase difference jumps discontinuously by $\Delta \theta$ at the flux bias $\Phi = \Phi_{0}/2$, changing from $\theta_0 $ to $ \theta_0 + \Delta \theta$ with $\Delta \theta \approx \pi$. This first-order transition is expected for typical skewness ($|S_{\rho}| \approx 0.2$) of our device.
%The jump occurs in such a way that the gauge-invariant phase difference $\gamma_{t,b} = \theta_{t,b} + e \int \mathbf{A} d\mathbf{l}  /\hbar $ on the weaker junction winds from $0$ to $2\pi$ as $\Phi$ is increased from 0 to $\Phi_0$, while the other junction remains close to $0$.
%
%For finite skewness and small asymmetry, as $\Phi$ is increased across $\Phi_{0}/2$, $\theta_{0}$ jumps discontinuously
%from $\theta_{0}=\delta\theta(\phi)\ll\pi$ to $\pi-\delta\theta(2\pi-\phi)$
%from $\theta_0=0$ to $\pi$ as  the magnetic
%flux is increased from $\Phi=0$ to $\Phi=\Phi_{0}/2$. 
%by $\theta_0 \to \theta_0 + \Delta \theta$ with $\Delta \theta \approx \pi$. 
%
%For an asymmetry larger than a critical value set by the skewness (or for zero skewness), 
%the phase winding is continuous.
%the equilibrium phase
%bias winds continuously from $\theta_{0} = 0$ to $\theta_{0} + \pi$ as the magnetic
%flux is increased from $\Phi=0$ to $\Phi=\Phi_{0}/2$.
We confirmed this result from calculations for the phenomenological nano-SQUID model, as shown in Fig.~S7 in the supplement.

%To fully substantiate the interpretation presented above, we performed a numerical simulation of a TI-nanowire Josephson junction using a tight-binding model (details presented in Methods). Comparison between the simulation results and the relevant experimental data for the critical currents $I_\mathrm{c}^+$ and $|I_\mathrm{c}^-|$, together with the diode efficiency, is shown in Fig.~\ref{fig:Fig6}. As the chemical potential is placed within the bulk band gap of the 3D TI, the current is carried through the top and bottom surfaces of the nanowire, modeling the nano-SQUID picture. The simulation reproduces both the behavior of critical currents as well as the magnitude of the diode effect faithfully, exhibiting sign changes of $\eta$ at half flux quanta and requiring the asymmetry between the top and bottom surface.

\section{Discussion}

Now we discuss the relevance of topology in the present experiment.
Since our TI-nanowire junction effectively forms a nano-SQUID consisting of two SNS junctions at the top and bottom TI surfaces, each SNS junction can be in the topological phase with %odd-parity Andreev bound states 
odd fermion parity in its ground state when the phase difference is between $\pi$ and $3\pi$ (mod $4\pi$) \cite{Fu2008}. 
Since we have two SNS junctions, the TI-nanowire junction as a whole is in the topological phase when only one of the two SNS junctions is topological \cite{Nikodem2024}; in this case, Majorana zero-modes are expected to show up at the ends of the TI nanowire. It was theoretically shown in Ref.  \citenum{Nikodem2024} that when the two junctions are asymmetric, this topological phase is realized in equilibrium in the magnetic-flux range of $(n-\frac{1}{2})\Phi_{0} < \Phi < (n+\frac{1}{2})\Phi_{0}$ with odd-integer $n$.
%When, on the other hand, the top and bottom junctions are both in the topological phase, the total parity of the system is restored to be even and the Majorana zero-modes will disappear.
%{\color{blue} For our TI-nanowire Josephson junction, the topological superconducting phase is realized when an odd number $n$ of flux quanta threads the junction, for the entire range of flux bias from $ (n - \frac{1}{2})\Phi_0$ to $(n + \frac{1}{2})\Phi_0$  \cite{Nikodem2024}. }

%We argue that the transition to the topological regime in equilibrium coincides with a sign reversal of the Josephson diode effect in our platform. Such topological phase transition corresponds to a fermion parity change of the $k_x = 0$ mode, which is the topological invariant of the one-dimensional topological superconductor  \cite{Kitaev2001}. This mode carries a large supercurrent because its dispersion must cross zero energy to facilitate the parity exchange as $\theta$ winds from 0 to $2\pi$. Changing the occupation of this mode reverses its supercurrent contribution and therefore it leads to a reversal of the dominant supercurrent direction. This phenomenology was also observed in the InAs/Al platform realizing topological high-transparency Josephson junctions \cite{Banerjee2023d}.

It is important to note that in this nano-SQUID, both the diode sign-reversal and the topological phase transition are pinned to half flux quanta by a common origin: the restoration of time-reversal symmetry at these magnetic fluxes. Time-reversal symmetry pins the phase differences of the top and bottom Josephson junctions to $0$ or $\pi$, where the latter is the topological phase transition point of a single junction \cite{Fu2008}. 
Importantly, the asymmetry between the top and bottom junctions can be controlled by the gate voltage $V_G$ in our setup; the junction having the smaller Josephson energy will experience the phase winding, thereby undergoing the topological transition. 
Upon exchanging the top-bottom asymmetry by gating, the junction that experiences the $0$--$\pi$ transition switches, which coincides with the gate-induced sign reversal of the diode effect. 

In the experiments to detect the Josephson diode effect, a bias current is passed through the junction and it dictates the phase difference across the junction according to the current-phase relation. This means that the phase is not a free parameter in the current-biased experiment and the persistence of the topological phase is not guaranteed upon current biasing. It is an interesting topic of future research to clarify the role of current bias to control the topological phase transition.
%Our simulations discussed in the previous section are performed under this condition, and it clarifies when the topological regime is realized. 
%However, when one would use the TI-nanowire junction for braiding experiments, there will be no current bias and only $B_{||}$ will be applied to bring the system into the topological regime. In such a case, the phase difference will become a free parameter which adjusts itself to minimize the free energy \cite{Pientka2017}. It would be very interesting to study the Andreev bound states and Majorana zero-modes in such a situation, possibly by using a tunnel contact \cite{Banerjee2023d}.

In conclusion, we found that the intrinsic nano-SQUID nature of the TI-nanowire junction \cite{Nikodem2024} leads to a large Josephson diode effect, whose size and sign can be tuned by the parallel magnetic field $B_{||}$ and the back-gate voltage $V_G$. Our theoretical modeling shows that the sign change in the Josephson diode efficiency $\eta$ as a function of $B_{||}$ is accompanied by a topological phase transition in equilibrium, making the TI-nanowire junction interesting not only for SC electronics but also for studying topological superconductivity and Majorana zero-modes.
Moreover, given that other tunable Josephson diodes based on SQUID geometry tend to be large and complicated \cite{Souto2022, Ciaccia2023}, the small and simple nature of our device is an advantage for integrating the Josephson diodes in large-scale circuits, although the necessity of a magnetic field can be a limiting factor.

\section{Materials and Methods}

\textbf{Materials and device fabrications:}
Bulk single crystals of BiSbTeSe$_2$ were grown using the modified Bridgman method, following the procedure outlined in Ref.  \citenum{Ren2011}, with high-purity (99.9999\%) Bi, Sb, Te, and Se as starting materials. Thin flakes of BiSbTeSe$_2$ were mechanically exfoliated from a bulk crystal and transferred onto a degenerately-doped Si wafer coated with a 290 nm SiO$_2$ layer, which acts as a dielectric for the back gate. Flakes suitable for device fabrication were identified under an optical microscope. Nanowires and Josephson junctions were patterned using electron beam lithography, with ZEP 520A resist exposed via a Raith PIONEER Two system. The BiSbTeSe$_2$ flakes were dry-etched into nanowires using Ar plasma. Subsequently, 45 nm of Nb were sputter-deposited to form the junctions. The device geometry was precisely characterized using scanning electron microscopy and atomic force microscopy after the completion of measurements. \\

\vspace{2mm}
\textbf{Measurements:}
Transport measurements were conducted at the base temperature ($\sim$30 mK) of our Oxford Instruments TRITON 300 dry dilution refrigerator. To minimize noise, the electrical lines were equipped with RC and copper-powder filters. We utilized a quasi-four-probe configuration to measure the differential resistance d$V$/d$I$ across the Josephson junction employing a standard low-frequency lock-in technique, where a small AC current of 5 nA was superimposed on a DC bias current. Note that whenever a four-terminal measurement was not possible due to defected leads, a quasi three-probe technique was applied instead (which was the case for device A). To apply magnetic fields, we employed a 6/1/1-T superconducting vector magnet.\\

\vspace{2mm}
\textbf{Simulations:}
To simulate the behavior of our devices, we use a three-dimensional lattice model with two superconducting side electrodes and a central topological insulator region. The topological insulator component of the device is described using a standard phenomenological 3D massive Dirac fermion model. The Hamiltonian of the model is given by:
\begin{equation}
\begin{split}
    H_\mathrm{TI}(\mathbf{k}) = A_1 k_{||} s_z \sigma_x + A_2 (k_x s_x + k_z s_y)\sigma_x \\+ (M - B_1 k_{||}^2 - B_2 (k_x^2 + k_z^2)) s_0 \sigma_z \\+ (C + D_1 k_{||}^2 + D_2 (k_x^2 + k_z^2) - \mu_\mathrm{TI}) s_0 \sigma_0
    % H_\mathrm{TI}(\mathbf{k}) = A_1 k_z s_z \sigma_x + A_2 (k_x s_x + k_y s_y)\sigma_x \\+ (M - B_1 k_z^2 - B_2 (k_x^2 + k_y^2)) s_0 \sigma_z \\+ (C + D_1 k_z^2 + D_2 (k_x^2 + k_y^2) - \mu_\mathrm{TI}) s_0 \sigma_0
\end{split}
\end{equation}
where $s_i$ and $\sigma_i$ are Pauli matrices describing spin and orbital degrees of freedom, respectively. The niobium superconductor that forms the Josephson junction is described using a simple parabolic band with the Hamiltonian:
\begin{equation}
    H_\mathrm{SC}(\mathbf{k}) = \left(t (k_x^2 + k_{||}^2 + k_z^2) - \mu_\mathrm{SC}\right)s_0 \sigma_0 
\end{equation}
Both Hamiltonians are discretized on a cubic lattice with the spacing $a_\mathrm{lat} = 1\,\mathrm{nm}$ using the finite difference method with nearest-neighbor hoppings. Since the devices under study are much longer ($L>$ 1 $\mu$m) than their width or height ($W, H$; order of 10\,nm), it is assumed that translational invariance is retained in direction parallel to the nanowire and thus $k_{||}$ remains a good quantum number.

After the discretization, both systems are combined in a single real-space Hamiltonian, which is given by:
\begin{equation}
    H_0(x) = H_\mathrm{SC} \theta(-x) + H_\mathrm{TI} (\theta(x) - \theta(W - x)) + H_\mathrm{SC} \theta(x-W)
\end{equation}
where $W$ is the width of the TI nanowire and $\theta(x)$ is the Heaviside step function. The matrix representation of this real-space Hamiltonian was constructed using Kwant package  \cite{Groth2014}. We include the effect of magnetic field via the Peierls' substitution, where the hoppings between sites $\mathbf{r}_n$ and $\mathbf{r}_m$ are replaced by:
\begin{equation}
    t_{mn} \rightarrow t_{mn} \exp\left(-i\tau_z\frac{e}{\hbar} \int_{\mathbf{r}_n}^{\mathbf{r}_m} d\mathbf{r}\cdot \mathbf{A}(\mathbf{r})\right)
\end{equation}
where $\tau_z=\pm1$ for particle and hole degrees of freedom, respectively. In all the calculations we use the gauge where $\mathbf{A}(\mathbf{r}) = -B_{||} z \hat{x}$. This assumes that the magnetic field penetrates the device fully, which is justified as the thickness of the superconducting electrodes (about 45 nm) is comparable to the London penetration depth of niobium. The flux through the junction is then established as $\Phi = B_{||} \tilde{A}$, where $\tilde{A}$ is the effective junction area.

To enable the SC diode effect, asymmetry between the top and bottom surface state of the nanowire has to be introduced through the gate electrostatic potential, which we model by a gradient of the chemical potential inside of the TI $\mu_\mathrm{TI}(z) = \mu_\mathrm{TI} + V_g z/H$.

The superconductivity then is treated using Bogoliubov-de Gennes (BdG) Hamiltonian:
\begin{equation}
    H_\mathrm{BdG} = \begin{pmatrix}
                        H_0(\mathbf{k}) & \Delta \\
                        \Delta^\dagger & -H_0^T(-\mathbf{k})
                     \end{pmatrix}
\end{equation}
where the superconducting order parameter $\Delta$ is:
\begin{equation}
    \Delta(x) = \Delta_0 \left(\exp(-i \phi / 2) \theta(-x) +  \exp(i \phi / 2) \theta(x - W)\right) is_y\sigma_0
\end{equation}
with $\phi$ being the phase difference between the superconducting condensates.

\begin{figure}
\centering
\includegraphics{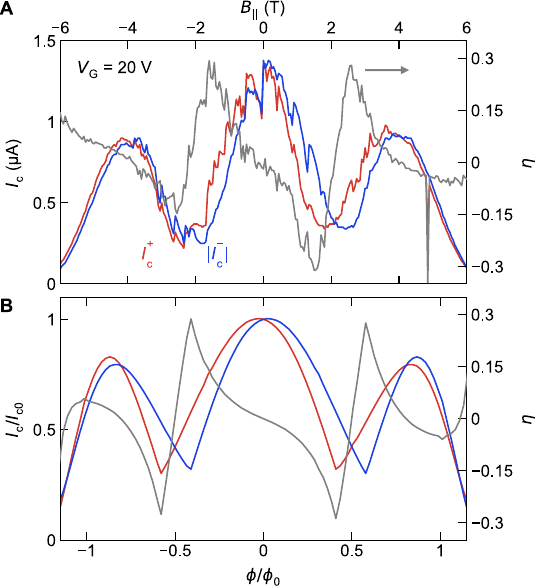}
\caption{\linespread{1.05}\selectfont{}
{\bf Comparison with a theoretical simulation. } 
(\textbf{A}) Experimental data of the critical currents $I_\mathrm{c}^+$ and $|I_\mathrm{c}^-|$ as a function of parallel magnetic field, together with the diode efficiency as measured in device A. (\textbf{B}) Simulation based on a tight-binding model of a TI nanowire proximitized from both sides by a conventional superconductor. This simulation result corresponds to the case of $S_{\rm t} = -0.2$, $S_{\rm b} = -0.2$, and $I_{0,\rm t}/I_{0,\rm b}$ = 0.8 in the phenomenological model, Eq.~\eqref{eq:CPR_rho}.}
\label{fig:Fig6}
\end{figure}

With such a setup constructed, we then calculate the current-phase relations for varying magnetic field by using the excitation spectrum of the BdG Hamiltonian. The Josephson current density across the quasi-1D nanowire is expressed through the positive eigenvalues that are a function of the momentum component parallel to the nanowire axis:
\begin{equation}
    j(\phi) = - \frac{e}{\hbar} \sum_{\epsilon_n > 0} \int \frac{dk_{||}}{2\pi} \tanh\left(\frac{\epsilon_n(k_{||})}{2 k_B T}\right) \frac{d\epsilon_n(k_{||})}{d\phi}
\end{equation}
where $k_B$ is the Boltzmann constant and $T$ is the temperature. The integration is performed by discretizing momenta and calculating the spectrum at 281 points for $k_{||} a_\mathrm{lat}$ in the range [-1.4, 1.4]. From the supercurrent density the total current through the junction is obtained as $I(\phi) = j(\phi) L$, and then the critical supercurrent in both directions is determined as $I_{c+} = \mathrm{max}(I(\phi))$ and $I_{c-} = \mathrm{min}(I(\phi))$, respectively. In the simulations the parameters are as follows: $a_\mathrm{lat}=1$, $A_1 =0.25$, $A_2=0.35$, $B_1=-0.3$, $B_2=-0.6$, $C=0$, $D_1=0.025$, $D_2=-0.05$, $M=-0.25$, $\mu_\mathrm{TI}=0.0$, $V_g=0.05$, $t=1.0$, $\mu_\mathrm{SC}=1.0$, $\Delta_0=0.08$, $W=14$, $W_\mathrm{SC}=16$, $H=15$, $L=1500$, with all the energies expressed in eV and all the lengths in nm. The phenomenological model parameters are chosen such that the bulk gap of the TI is in agreement with the experimentally determined gap $250-300$ meV of BiSbSeTe$_2$. These parameters also translate to a topological surface state with a Fermi velocity $v_F \approx 4\times10^5\,$m/s. Since the chemical potential changes between the top and the bottom surfaces of the TI, the Fermi wave length changes as well, but generally it is between 25 and 65 nm. The imperfect transmission at the superconductor-3DTI interface was modeled by the Fermi wave vector mismatch, which was $k_\mathrm{F,SC}/k_\mathrm{F,TI}\approx 4$. To make the simulations numerically tractable the use of lattice discretization with 1 nm spacing required increasing the superconducting gap magnitude from the experimental value of $\Delta_\mathrm{exp} \approx 0.9$ meV. The TI-nanowire width is chosen so that the mini-gap of the calculated spectrum of states in the junction is equal to about 40 meV, which is approximately 1/2 of the gap of the SC electrodes in the model ($\Delta_0 = 80$ meV). This fraction was chosen to be consistent with the size of the induced gap measured in the studied device.
One can also estimate the fictituous superconducting coherence length in the simulation to be $\xi = \hbar v_F/\Delta_0=3.3\,$nm using $v_F$ of the TI surface state and the model SC gap in the leads. The $V_g$ value used for the simulation in Fig.~\ref{fig:Fig6}B is chosen because it produces the largest diode efficiency.

%%%%%%%%%%%%%%%% MAIN TEXT FIGURES %%%%%%%%%%%%%%%

%%%%%%%%%%%%%%%% MAIN TEXT TABLES %%%%%%%%%%%%%%%

%%%%%%%%%%%%%%%% REFERENCES %%%%%%%%%%%%%%%

\clearpage % Clear all remaining figures and tables then start a new page

% The list of references goes after the main text and before the acknowledgements
% When preparing an initial submission, we recommend you use BibTeX, like this:
%
\bibliography{Bibliography} % for a file named science_template.bib
\bibliographystyle{sciencemag}

% After the paper has completed peer review and been revised ready for acceptance,
% you should comment out the lines above and copy-paste the contents of your .bbl
% file here instead. This will help ensure that our conversion software works correctly.
% Remember to re-run BibTeX first - check the timestamp!
%
% Example of the first three entries copy-pasted from science_template.bbl:
%
%\begin{thebibliography}{1}
%
%\bibitem{example}
%A.~N. {Author}, An example reference. \emph{Journal of Improbable Research}
%  \textbf{1}, 67 (2020).
%
%\bibitem{example2}
%F.~M. {Surname}, S.~{Author}, A second example. \emph{Interesting Research
%  Letters} \textbf{32}, 897 (2019).
%
%\bibitem{example_preprint}
%P.~{One}, P.~{Two}, P.~{Three}, {An unpublished preprint}. \emph{preprint}
%  (2021), arXiv:2101.12345.
%
%\end{thebibliography}

%%%%%%%%%%%%%%%% ACKNOWLEDGEMENTS %%%%%%%%%%%%%%%
%\section*{Acknowledgments:}
\paragraph*{Funding:}
This project has received funding from the European Research Council (ERC) under the European Union’s Horizon 2020 research and innovation program (Grant Agreement No. 741121) and was also funded by the Deutsche Forschungsgemeinschaft (DFG, German Research Foundation) under Germany’s Excellence Strategy - Cluster of Excellence Matter and Light for Quantum Computing (ML4Q) EXC 2004/1 - 390534769, as well as by the DFG under CRC 1238 - 277146847 (Subprojects A04 and B01). EN acknowledges support by the Studienstiftung des deutschen Volkes. This work was completed in part with resources provided by the Research Computing Data Core at the University of Houston.
MG acknowledges support from the German Research
Foundation under the Walter Benjamin program (Grant
Agreement No. 526129603).
% and by the Air Force Office of Scientific Research under award number FA2386-24-1-4043.
LF is partly supported by a Simons Investigator Award from the Simons Foundation.
\paragraph*{Author contributions:} 
YA and JS conceived the project. MB grew the TI crystals. JS and EN fabricated the devices, performed the experiments and analysed the data with inputs from JF and YA. MG, MP, HFL, and LF performed the theoretical analyses. EN, YA, MG, MP, HFL, and LF wrote the manuscript with input from all authors.
\paragraph*{Competing Interests:} 
The authors declare they have no competing interests.
\paragraph*{Data and code availability:}
The data and codes that support the findings of this study are available at the online depository zenodo with the identifier {10.5281/zenodo.14576331} and Supplementary Information.

\end{document}

\begin{center}
\section*{Supplementary materials for \\``Tunable superconducting diode effect in a topological nano-SQUID''}

	Ella~Nikodem$^{1\dagger}$,
    Jakob~Schluck$^{1\dagger}$,
	Max~Geier$^{2}$,
    Micha\l ~Papaj$^{3}$,
	Henry~F.~Legg$^{4,5}$,\\
    Junya~Feng$^{1}$,
    Mahasweta~Bagchi$^{1}$,
    Liang~Fu$^{2}$,
    Yoichi~Ando$^{1\ast}$\\
	% Additional lines of authors should be inserted using the \and command (not \\)
	% Institution list, in a slightly smaller font
	\small$^{1}$Physics Institute II, University of Cologne, Zülpicher Str. 77, 50937 Köln, Germany.\\
	\small$^{2}$Department of Physics, Massachusetts Institute of Technology, Cambridge MA 02139, USA.\\
    \small$^{3}$Department of Physics and Texas Center for Superconductivity at the University of Houston (TcSUH),\\ \small Houston, TX 77204, USA.\\
    \small$^{4}$Department of Physics, University of Basel, Klingelbergstrasse 82, 4056 Basel, Switzerland.\\
    \small$^{5}${SUPA, School of Physics and Astronomy, University of St Andrews,\\ \small North Haugh, St Andrews, KY16 9SS, United Kingdom.}\\
	% Identify at least one corresponding author, with contact email address
	\small$^\ast$Corresponding author. Email: ando@ph2.uni-koeln.de\\
	% Joint contributions can be indicated like this
	\small$^\dagger$These authors contributed equally to this work.
\end{center}

% Fill out the numbers for each type of supplementary material,
% and delete any lines that aren't applicable.
% These are just example numbers that don't match the rest of this template.
\subsubsection*{This PDF file includes:}
Supplementary Text\\
Figures S1 to S8\\
Table S1\\
\newpage

%%%%%%%%%%%%%%%% SUPPLEMENTARY TEXT %%%%%%%%%%%%%%%

\subsection*{Supplementary Text}
\subsection*{Experimental details and additional data}

\subsubsection*{Sample geometry}
%\vspace{-5mm}

\begin{figure}[h]
\centering
\includegraphics[width=\textwidth]{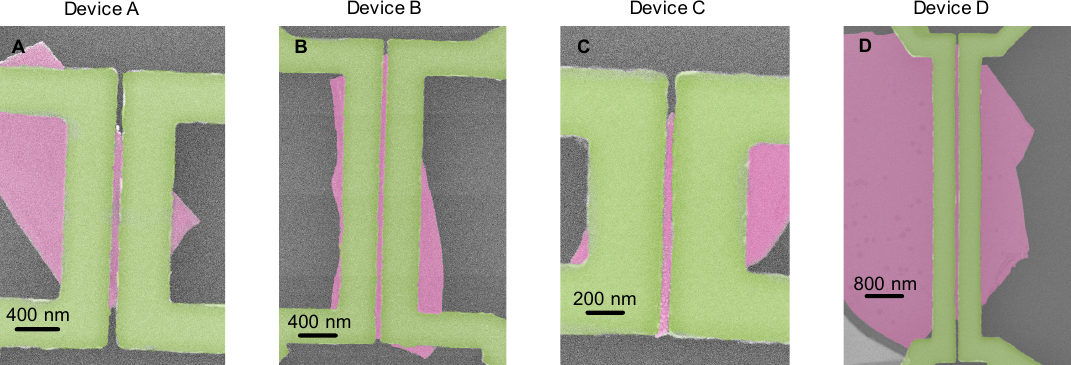}
\caption{\linespread{1.05}\selectfont{}
\textbf{SEM images of devices A-D.}}
\label{fig:S1}
\end{figure}

Following the transport measurements, the device geometry was accurately characterized using the scanning electron microscopy (SEM) images shown in Figs. \ref{fig:S1}A-D to determine the length $L$ and width $W$ of the nanowires. The nanowire thickness was measured using atomic force microscopy (AFM). The geometric parameters for devices A, B, C, and D are summarized in Table \ref{table:Tab1}.

% \normalsize

\subsubsection*{Magnetic field alignment}
%\vspace{-5mm}

The precise alignment of the magnetic field relative to the junction's coordinate system is crucial for studying the junction properties in magnetic fields parallel to the nanowire axis, $B_\parallel$. Misalignment introduces a nonzero $B_z$ component, leading to phase winding and a current modulation along the junction, as observed in a Fraunhofer measurement. To achieve good alignment, multiple Fraunhofer-type measurements were performed with intentionally-applied small $B_z$ fields in the presence of large nominal $B_\parallel$ fields. These data reveal a systematic shift in the position of the maximum critical current $I_c$ arising from a spurious $B_z$ component caused by misalignment. By identifying the necessary $B_z$ value to compensate the spurious $B_z$ component for each nominal $B_\parallel$, we achieved the necessary alignment of $B_\parallel$. The precise procedure is described in Ref. [24].

On the other hand, a small misalignment of the magnetic field within the device plane from the nanowire axis does not affect the data in any noticeable manner.

\subsubsection*{Reproducibility of the Josephson diode effect}
%\vspace{-5mm}

\begin{figure}[h]
\centering
\includegraphics[width=\textwidth]{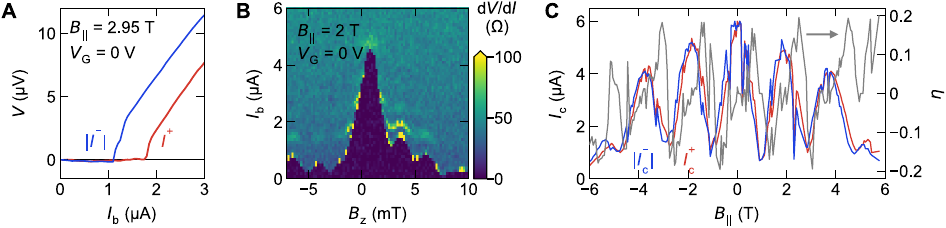}
\caption{\linespread{1.05}\selectfont{}
\textbf{Reproducibility in device B.} \textbf{A,} $I$-$V$ characteristics at $B_\parallel=2.95$\,T and $V_\mathrm{G} = 0$\,V. The negative bias curve (blue) was flipped horizontally and vertically. The current sweep was always from zero. \textbf{B,} Color map of d$V$/d$I$ as a function of $B_z$ and $I_\mathrm{b}$ at $B_\parallel=2$\,T and $V_\mathrm{G} = 0$\,V. \textbf{C,} $I_c$ for positive bias currents (red) and negative bias currents (blue), as well as the diode efficiency $\eta$ (grey) as a function of $B_\parallel$ at $V_\mathrm{G} = 0$\,V. This device was not gate-tunable. Measurements were at $T$ = 30 mK.}
\label{fig:S2}
\end{figure}

\begin{figure}[h]
\centering
\includegraphics[width=\textwidth]{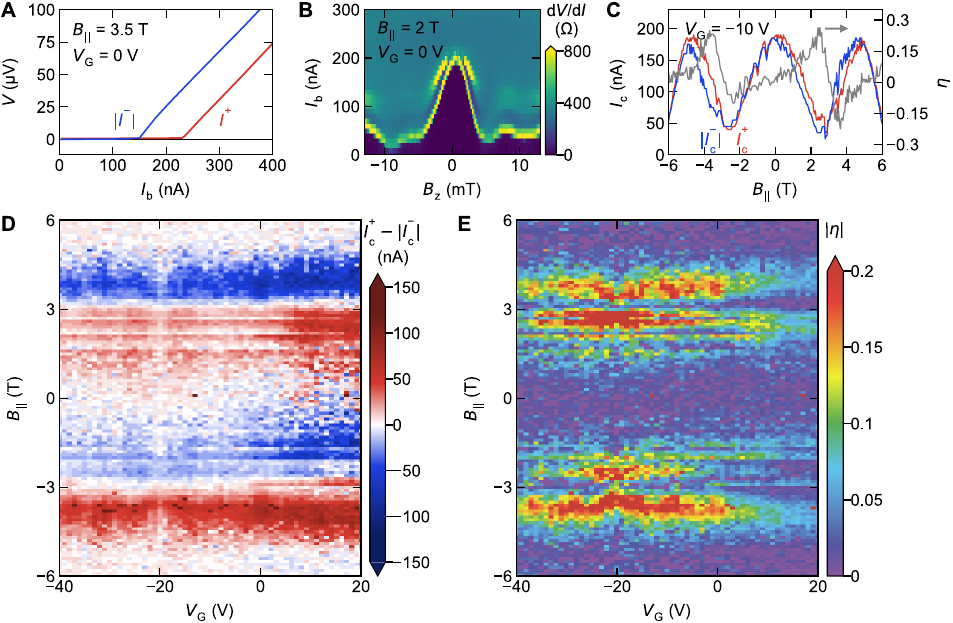}
\caption{\linespread{1.05}\selectfont{}
\textbf{Reproducibility in device C.} \textbf{A,} $I$-$V$ characteristics at $B_\parallel=3.5$\,T and $V_\mathrm{G} = 0$\,V. The negative bias curve (blue) was flipped horizontally and vertically. The current sweep was always from zero. \textbf{B,} Color map of d$V$/d$I$ as a function of $B_z$ and $I_\mathrm{b}$ at $B_\parallel=2$\,T and $V_\mathrm{G} = 0$\,V. \textbf{C,} $I_c$ for positive bias currents (red) and negative bias currents (blue), as well as the diode efficiency $\eta$ (grey) as a function of $B_\parallel$ at $V_\mathrm{G} = -10$\,V. \textbf{D,} Color mapping of $I_\mathrm{c}^+-|I_\mathrm{c}^-|$ as a function of $V_\mathrm{G}$ and $B_\parallel$. \textbf{E,} Color mapping of $|\eta|$ as a function of $V_\mathrm{G}$ and $B_\parallel$. Measurements were at $T$ = 30 mK.}
\label{fig:S3}
\end{figure}

\begin{figure}[h]
\centering
\includegraphics[width=\textwidth]{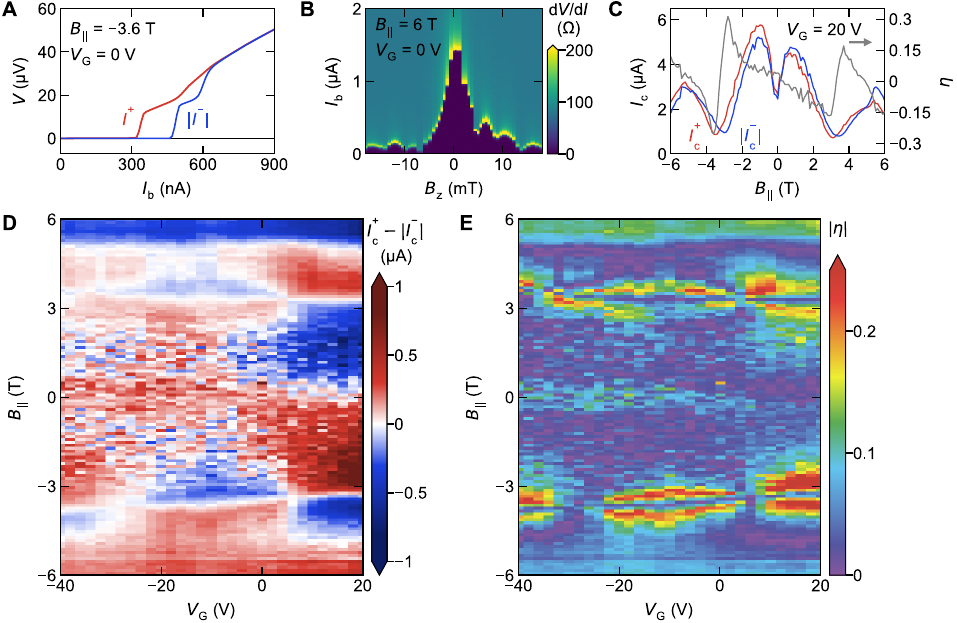}
\caption{\linespread{1.05}\selectfont{}
\textbf{Reproducibility in device D.} \textbf{A,} $I$-$V$ characteristics at $B_\parallel=-3.6$\,T and $V_\mathrm{G} = 0$\,V. The negative bias curve (blue) was flipped horizontally and vertically. The current sweep was always from zero. \textbf{B,} Color map of d$V$/d$I$ as a function of $B_z$ and $I_\mathrm{b}$ at $B_\parallel=6$\,T and $V_\mathrm{G} = 0$\,V. \textbf{C,} $I_c$ for positive bias currents (red) and negative bias currents (blue), as well as the diode efficiency $\eta$ (grey) as a function of $B_\parallel$ at $V_\mathrm{G} = 20$\,V. \textbf{D,} Color mapping of $I_\mathrm{c}^+-|I_\mathrm{c}^-|$ as a function of $V_\mathrm{G}$ and $B_\parallel$. \textbf{E,} Color mapping of $|\eta|$ as a function of $V_\mathrm{G}$ and $B_\parallel$. Measurements were at $T$ = 30 mK.}
\label{fig:S4}
\end{figure}

In Figs. \ref{fig:S2}, \ref{fig:S3}, \ref{fig:S4}, we show data for three more devices B, C, and D not presented  in the main text. As shown in Fig. \ref{fig:S1}, the device design is essentially the same as the one discussed in the main text. All key features are reproduced in these devices. In the presence of a finite $B_\parallel$, all of the devices develop a pronounced Josephson diode effect that is periodic in $B_\parallel$. Note that in contrast to devices C and D, device B was not gate-tunable due to an accidental short to the back gate. Intriguingly, in device D we observe two sign changes of the diode efficiency as a function of $V_\mathrm{G}$ as shown in Fig. \ref{fig:S4}D.

\subsubsection*{Josephson diode effect at 1 K}
%\vspace{-5mm}

\begin{figure}[h]
\centering
\includegraphics[width=\textwidth]{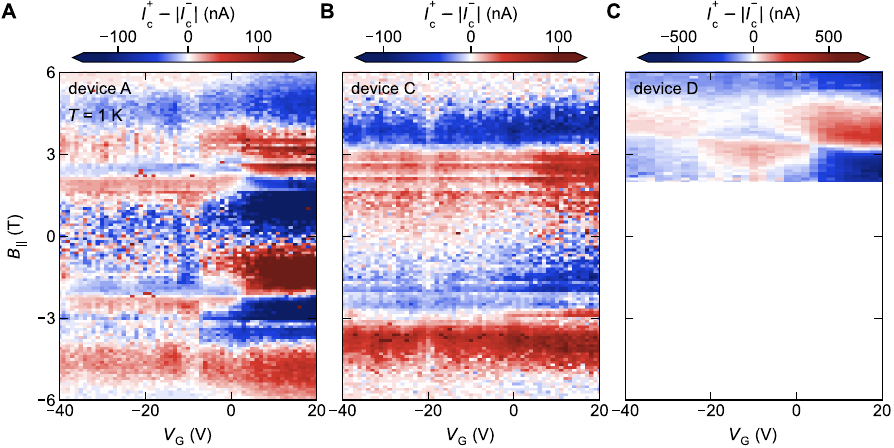}
\caption{\linespread{1.05}\selectfont{}
\textbf{A, B, C,} Color mapping of $I_\mathrm{c}^+-|I_\mathrm{c}^-|$ at $T=1$\,K as a function of $V_\mathrm{G}$ and $B_\parallel$ in devices A, C, and D. The measurement of Device D at 1 K was performed only for a limited range of $B_\parallel$, since there was little change from the behavior at 30 mK shown in Fig. \ref{fig:S4}.}
\label{fig:S5}
\end{figure}

\begin{figure}[h]
\centering
\includegraphics[width=\textwidth]{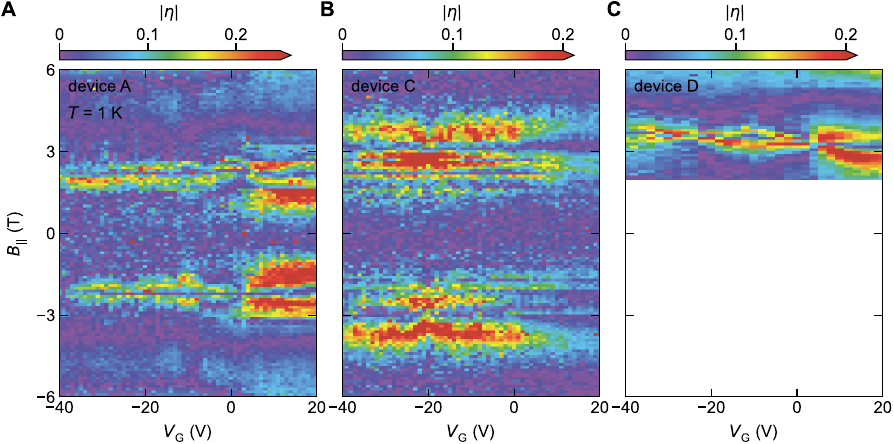}
\caption{\linespread{1.05}\selectfont{}
\textbf{A, B, C,} Color mapping of $|\eta|$ at $T=1$\,K as a function of $V_\mathrm{G}$ and $B_\parallel$ in devices A, C, and D.}
\label{fig:S6}
\end{figure}

Figures~\ref{fig:S5} and \ref{fig:S6} show $I_\mathrm{c}^+-|I_\mathrm{c}^-|$ and the diode efficiency $|\eta|$, respectively, as a function of $V_\mathrm{G}$ and $B_\parallel$ for devices A, C, and D measured at $T=1$\,K. The tunability of $I_\mathrm{c}^+-|I_\mathrm{c}^-|$ remains unaffected by the increased temperature, while the magnitude of $I_c$ in both bias directions is slightly reduced.
The diode efficiency is largely robust against the temperature increase.

\vspace{-5mm}

\subsection*{Mechanism of the sign reversal of the diode effect}

\begin{figure}
\centering
\includegraphics[width=0.4\linewidth]{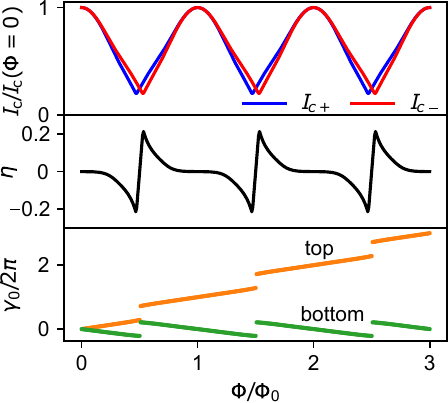}
\caption{\textbf{\label{fig:Fig5} Phenomenological model}.
Critical currents in positive and negative directions $I_{{\rm c}\pm}$,
diode efficiency $\eta$, and gauge-invariant equilibrium phase bias $\gamma_{0, {\rm t/b}} = \theta_{0} + \int \mathbf{A}{\rm d}\mathbf{l} = \theta_{0} \pm \pi \Phi/\Phi_0$ at top- and bottom surfaces
as a function of magnetic flux. Shown here are the model calculations from Eq.~(1) in the main text with skewness $S_{{\rm t}}=S_{{\rm b}}=-0.2$ and asymmetry $I_{{\rm 0,t}}/I_{{\rm 0,b}}=0.8$. }
\end{figure}

\begin{figure}
    \centering
    \includegraphics[width=0.45\linewidth]{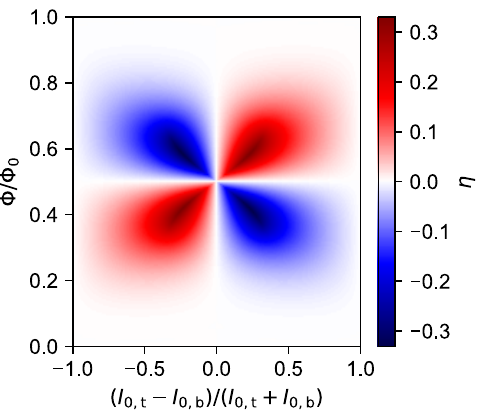}
    \caption{Diode efficiency calculated from Eq.~(1) in the main text, as a function of $\Phi$ and top-bottom asymmetry $\frac{I_{\rm 0,t} - I_{\rm 0,b}}{I_{\rm 0,t} + I_{\rm 0,b}}$. The skewness parameter setting the weight of the second harmonic is $S_{\rm t} = S_{\rm b} = -0.2$. }
    \label{fig:S-theo-eta}
\end{figure}

The phenomenological nano-SQUID model, Eq.~(1) in the main text, predicts a sign reversal of the diode efficiency at integer multiples of half a magnetic flux quantum, see Fig.~\ref{fig:Fig5} for results derived from the nano-SQUID model. This is a robust feature of the nano-SQUID description because it is tied to the restoration of time-reversal symmetry at these flux values. 
Notably, the sign reversal of the diode efficiency is also observed
in the experimental data (Fig. 2A in the main text) for the whole range of gate
voltages at the magnetic field corresponding to the flux $\Phi=\Phi_{0}/2$.
Qualitative agreement of the experimental data with the phenomenological
model supports the conclusion that the supercurrent is predominantly carried by the
top and bottom surface states separately.
%while direct tunneling between the top and bottom surfaces is suppressed. Importantly, the absence of $h/e$-periodic features indicates the suppression of coherent tunneling between the top and bottom surfaces via the surface states of the side surfaces, which are gapped by proximity to the superconductor. 
Further, we compute the diode efficiency as a function of top-bottom asymmetry $\frac{I_{\rm 0,t} - I_{\rm 0,b}}{I_{\rm 0,t} + I_{\rm 0,b}}$ and magnetic flux, with the result shown in Fig.~\ref{fig:S-theo-eta}. The diode efficiency changes sign around the time-reversal symmetry points $\Phi = \frac{n}{2}\Phi_0,\, n \in \mathbb{Z}$ as well as when the top-bottom asymmetry is exchanged. Already with fixed skewness $S_{\rm t} = S_{\rm b} = -0.2$, the diode efficiency approaches the theoretical maximum of $|\eta| = \frac{1}{3}$ that is achievable with current-phase relations containing only first- and second harmonic components. (The theoretical maximum  $|\eta| = \frac{1}{3}$ for current-phase relations with first- and second-harmonic components of the form $I_0 \left( \sin(\theta) + S \sin(2 \theta + \delta \theta) \right)$ is reached for $S = 0.5$ and $\delta \theta = \pi/2$.)

The sign reversal of the diode effect as a function of the gate voltage $V_{G}$ also appears near $V_{G} \approx$ 0 in the experimental data (Fig.~2E) for several magnetic field values.  The direction of the diode effect can thus be controlled by the gate voltage $V_{G}$, in addition to the magnetic flux that threads the nanowire. 
Furthermore, for a large positive gate voltage $V_{G}$, the experimental diode efficiency approaches the theoretical maximum of $|\eta| = \frac{1}{3}$ mentioned above. 
While the experimental current-phase relations may contain any order of harmonics, we expect that the second harmonic dominates over the higher harmonics because the long nanowire has many transverse modes for which higher harmonics are suppressed by specular reflection competing with the Andreev reflection at the superconductor interface. This picture is also confirmed by our microscopic simulations, where we observed $|\eta| < 1/3$.

%%%%%%%%%%%%%%%% SUPPLEMENTARY FIGURES %%%%%%%%%%%%%%%

\clearpage

\clearpage

\clearpage

\clearpage

\clearpage

\clearpage

\clearpage

\clearpage

\clearpage
\begin{table}[h]
\small
\centering 
    \begin{tabular}{ clclclcl}
    Device & $d$ (nm) & $W$ (nm) & $L$ (\textmu m) \\ 
    \hline
    A & 15 & 60 & 1.2\\ 
    B & 22 & 80 & 2.4 \\
    C & 13 & 60  & 1.5\\  
    D & 12 & 70 & 4.9 \\
    \end{tabular}
    \caption{\textbf{Nanowire dimensions for devices A--D.}}
    \label{table:Tab1}
 \end{table}
 \normalsize

%%%%%%%%%%%%%%%% SUPPLEMENTARY REFERENCES %%%%%%%%%%%%%%%

% Do NOT include a reference list in the supplement.
% All references must be in a single list at the end of the main text.
% The copyeditors will ensure that the correct reference list appears with each version of the paper
% (print, HTML, PDF, mobile app, metadata for bibliographic databases etc.)